# The World-Wide Telescope, an Archetype for Online Science

Jim Gray, Microsoft Research
Alex Szalay, Johns Hopkins University

# The World-Wide Telescope, an Archetype for Online Science

Jim Gray
Microsoft Research
Gray@Microsoft.com

Alex Szalay
The Johns Hopkins University
Szalay@jhu.edu

**Abstract**: Most scientific data will never be directly examined by scientists; rather it will be put into online databases where it will be analyzed and summarized by computer programs. Scientists increasingly see their instruments through online scientific archives and analysis tools, rather than examining the raw data. Today this analysis is primarily driven by scientists asking queries, but scientific archives are becoming active databases that self-organize and recognize interesting and anomalous facts as data arrives. In some fields, data from many different archives can be cross-correlated to produce new insights. Astronomy presents an excellent example of these trends; and, federating Astronomy archives presents interesting challenges for computer scientists.

**Introduction**

Computational Science is a new branch of most disciplines. A thousand years ago, science was primarily *empirical*. Over the last 500 years each discipline has grown a *theoretical* component. Theoretical models often motivate experiments and generalize our understanding. Today most disciplines have both empirical and theoretical branches. In the last 50 years, most disciplines have grown a third, *computational* branch (e.g. empirical, theoretical, and computational ecology, or physics, or linguistics.)

Computational Science traditionally meant simulation. It grew out of our inability to find closed-form solutions for complex mathematical models. Computers can simulate these complex models.

Computational Science has been evolving to include information management. Scientists are faced with mountains of data that stem from four converging trends: (1) the flood of data from new scientific instruments driven by Moore's Law -- doubling their data output every year or so; (2) the flood of data from simulations; (3) the ability to economically store petabytes of data online; and (4) the Internet and computing Grid that makes all these archives accessible to anyone anywhere.

Scientific information management poses some Computer Science challenges. Acquisition, organization, query, and visualization tasks scale almost linearly with data volumes. By using parallelism, these problems can be solved within fixed times (minutes or hours).

In contrast, most statistical analysis and data mining algorithms are nonlinear. Many tasks involve computing statistics among sets of data points in a metric space. Pair-algorithms on $N$ points scale as $N^2$. If the data increase a thousand fold, the work and time can grow by a factor of a million. Many clustering algorithms scale even worse. These algorithms are infeasible for terabyte-scale datasets.

The new online science needs new data mining algorithms that use near-linear processing, storage, and bandwidth, and that can be executed in parallel. Unlike current algorithms that give exact answers, these algorithms will likely be heuristic and give approximate answers [Connolly, Szapudi].

**Astronomy as an Archetype for Online Science**

Astronomy exemplifies these phenomena. For thousands of years astronomy was primary empirical with few theoretical models. Theoretical astronomy began with Kepler is now co-equal with observation. Astronomy was early to adopt computational techniques to model stellar and galactic formation and celestial mechanics. Today, simulation is an important part of the field – producing new science, and solidifying our grasp of existing theories.

Astronomers are building telescopes that produce terabytes of data each year -- soon terabytes per night. In the old days, astronomers could carefully analyze each photographic plate. Now humans would take years just to analyze a single evening's observation. Rather, the data is fed to software pipelines that use massive parallelism to analyze the images, recognize objects, classify them, and build catalogs of these objects. Astronomers used data analysis tools to explore and visualize the data catalogs. Only when the astronomer sees something anomalous does she go back to the source pixels – hence most source data is never directly examined by humans.

Astronomy data is collected by dedicated instruments around the world. Each instrument measures the light intensity (flux) in certain spectral bands. Using this information, astronomers can extract hundreds of object attributes including the magnitude, extent, probable structure and morphology. Even more can be learned by combining observations from various times and from different instruments.

Temporal and multi-spectral studies require integrating data from several archives. Until recently, this was very difficult. But, with the advent of the high-speed Internet and with inexpensive online storage, it is now possible for scientists to compare data from multiple archives for analysis. Figure 2 shows one example of the importance of comparing temporal and multi-spectral information about an object.

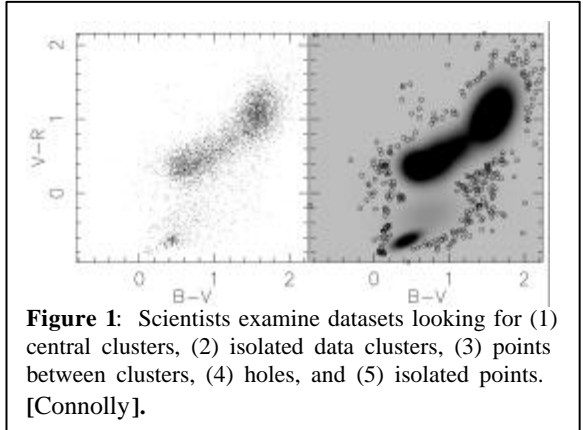

**Figure 1**: Scientists examine datasets looking for (1) central clusters, (2) isolated data clusters, (3) points between clusters, (4) holes, and (5) isolated points. [Connolly].

Astronomers typically get archival data by requesting large parts of the archive on magnetic tape, or smaller parts by File Transfer Protocol (FTP) over the Internet. The data arrives encoded as FITS files with a unique coordinate system and measurement regime [FITS]. The first task our astronomer faces is converting the "foreign" data into a "domestic" format and measurement system. Just as each computer scientist has dozens of definitions for the word "process", astronomy subgroups each have their own vocabulary and style. The scientist then analyzes the data with a combination of a scripting language (tcl and Python are popular) and a personal analysis toolkit acquired during their careers. Using these tools, the astronomer "greps" the data, applies statistical tests, and looks for common trends or for outliers. Figure 1 describes the general patterns that astronomers look for. The Astronomer uses visualization packages to "see" the data as 2D and 3D scatter plots.

This FTP-GREP metaphor does not work for terabyte-sized datasets. You can GREP or FTP a gigabyte in a minute.

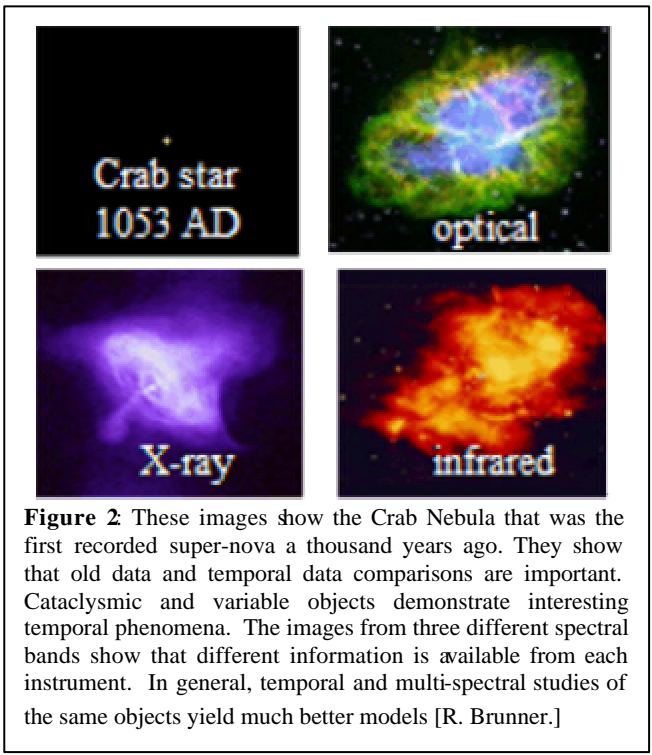

**Figure 2**: These images show the Crab Nebula that was the first recorded super-nova a thousand years ago. They show that old data and temporal data comparisons are important. Cataclysmic and variable objects demonstrate interesting temporal phenomena. The images from three different spectral bands show that different information is available from each instrument. In general, temporal and multi-spectral studies of the same objects yield much better models [R. Brunner.]

But it can take a day or more to GREP or FTP a terabyte, and scanning a petabyte would take several years. So, the old tools will not work in the future.

Database query systems have automatic indexing and parallel search that let people explore huge databases. A 100 Terabyte database occupies several thousand disks. Searching them one-at-a-time would take months, but a parallel search takes only an hour. More importantly, indices can focus the search to run in seconds or minutes. But, the datamining algorithms are super-linear, so new linear-time approximations that run on parallel computes are needed.

**Astronomy Data Will All be Online**
Nearly all the "old" astronomy data is online today as FITS files that can be FTPed to your site. Astronomers have a tradition of publishing their raw data after they have validated and analyzed it. We estimate that about half of the world's astronomical data is online today – about 100 terabytes in all.

Palomar Observatory did a detailed optical sky survey in the 1950's using photographic plates. Originally it was published via prints on plastic film, for about $25K per copy. That data has now been digitized and is on the Internet for free.

Several new all-sky surveys started in the 1990's. Together they will gather deep and statistically uniform surveys of the sky in about 20 spectral bands. Each survey will generate terabytes of data. They have large sky coverage, sound statistical plans, are well-documented, and are being designed to be federated into a *Virtual Observatory* [VO]. Software to process the data, to present the catalogs to the public, and to federate with other archives is a major part of each of these surveys – often comprising more than 25% of the project's budget.

In addition, the Astronomy community has been cross-correlating the scientific literature with the archives [NED, VIzieR]. Today, these tools allow you to point at an object and quickly find all literature about it and all the other archives with information on the object.

**The World-Wide Telescope**
The *World-Wide Telescope* (WWT) will emerge from the world's online astronomy data. It will have observations in all the observed spectral bands, from the best instruments back to the beginning of history. The "seeing" will always

good – the Sun, the Moon, and the clouds will not create dead-time when you cannot observe. Furthermore, all this data can be cross-indexed with the online literature. Today, you can find and study all the astronomy literature online with Google™ and AstroPh. In the future you should be able to find and analyze the underlying observational data just as easily.

The World-Wide Telescope will have a democratizing effect on astronomy. Professional and amateur astronomers will have nearly equal access to the data. The major difference will be that some will have much better data analysis tools and skills than others. Often, following up on a conjecture requires a careful look at the object using an instrument like the Hubble-Space Telescope, so there will still be many projects for those with privileged access to those instruments. But, for studies that analyze the global structure of the universe, tools to mine the online data will be a wonderful telescope in their own right.

The World-Wide Telescope will also be an extraordinary tool for teaching Astronomy. It gives students at every grade level access to the world's best telescope. It may also be a great way to teach Computational Science skills, because the data is real and well-documented, and has a strong visual component.

Building the World-Wide Telescope will require skills from many disciplines. First, it will require the active participation of the many Astronomy groups that gather data and produce catalogs. That is at least 75% of the effort – but once done, the challenge remains to make the data accessible. Making the data useful requires three additional components: (1) Good *database plumbing* to store, organize, query, and access the data – both in huge single-site archives, and in the distributed Internet database involving all the archives. This involves data management and just as importantly, meta-data management to integrate the different archives. (2) D*ata mining algorithms* and *tools* that can recognize and categorize data anomalies and trends. These algorithms will draw heavily on techniques from statistics and machine learning, but will also require new approaches that scale linearly with data size. Most of these algorithms will be generic, but some will require a deep understanding of Astronomy. (3) Finally, good *data visualization tools* are needed that make it easy to pose questions in a visual way and to "see" the answers.

The obvious challenges are the huge databases. Each spectral band is a few tens of terabytes. The multi-spectral and temporal dimensions grow data volumes to petabytes. So, automatic parallel search and good indexing technologies will be absolutely essential. But, we expect the large database search and index problems will be solved. There has been huge progress in the past, and more on the horizon.

In contrast, the problem of integrating heterogeneous data schemas has eluded solution for decades, and is now even more pressing. Automatically combining data from multiple data sources, each with its own data lineage, data units, data quality, and data conventions is a huge challenge. Today it is done painstakingly, one item at a time. The WWT must make it easy for Astronomers to publish their data on the Internet in understandable forms. The WWT must also make it easy for their colleagues to find and analyze this data using standard tools.

**The Virtual Observatory and SkyServer**
The SkyServer (http://SkyServer.SDSS.org/) is a prototype online telescope. It is just a small part of a larger *Virtual Observatory* being jointly built by the international astronomy community [VO]. SkyServer began as an attempt to make the Sloan Digital Sky Survey (SDSS) data easily available. The project expanded to include tools for datamining, an educational component, and federating the SDSS archives with others and with the literature.

The SkyServer gives interactive access to the data via a point-and-click virtual telescope view of the pixel data and via pre-canned reports generated from the online catalogs. It also allows ad hoc catalog queries. All the data is accessible via standard browsers. A Java™ GUI client interface that lets users pose SQL queries, while Python and EMACS interfaces allow client scripts to access the database. All these clients use the same public http/soap/xml interfaces.

The SkyServer database was designed to answer 20 queries that typify the kinds of questions an astronomer might ask of an archive [Szalay.] A typical query is "Find gravitational lens candidates." or "Create a grided count of galaxies satisfying a color cut." We were delighted to find that all the queries had fairly short SQL equivalents. Indeed, most were a single SQL statement. The queries all run in less than ten minutes and most run in less than a minute [Gray].

An anecdote conveys a sense of how the SkyServers' interactive access to the data can change the way Astronomers work. A colleague challenged us to find "fast moving" asteroids. This was an excellent test case – our colleague had written a 12 page Tcl script that had run for 3 days on the flat files of the dataset. So we had a benchmark to work against. It took a long day to debug our understanding of the data and to develop an 18-line SQL query. The resulting query runs in a minute. This interactive access (not 3-day access) allowed us to "play" with the data and find other objects. Being able to pose questions in a few hours and get answers in a few minutes changes the way one views the data: you can experiment interactively. When queries take three days and hundreds of lines of code, one asks many fewer questions and so gets fewer answers.

This and similar experiences convince us that interactive access to scientific data and datamining tools can have a huge impact on scientists' productivity.

The SkyServer is also an educational tool. Several interactive astronomy projects, from elementary to graduate level have been developed (in three languages). Interest in this aspect of the SkyServer continues to grow.

The SDSS data is public. Computer scientists have started using it in datamining and visualization research. Indeed we have a .1% edition that is about 1GB, a 5% edition that is about 100GB and the 100% edition will be about 10TB when complete. The 5% edition can be cloned for about 5,000$.

In parallel, colleagues at Caltech built http://VirtualSky.org that puts most of the Digital Palomar Sky Survey data online.

Having built web servers that provide HTML access to the data, the next step is to federate them into an international database with transparent access to the data. The datasets are already federated with the literature, you can point at an object and find everything written about it and find all other datasets that catalog that object [NED, VIZieR].

SkyQuery (http://SkyQuery.net) gives a taste of such a VO federation. Using web services technologies, it federates the optical SDSS archive at Fermi Lab with a radio survey [FIRST] archive at Johns Hopkins, and a Two Meter All Sky Survey (2MASS) archive at Cal Tech. Given a query, the *SkyQuery* portal queries these *SkyNodes* and uses them to *cross-correlate objects*. The query can be stated as: "For a certain class of objects find all information about matching objects in the other surveys." It combines this information and renders it in a graphic composed by a fourth web service. Automatically answering this query requires a uniform naming, coordinate system, measurement units, and error handling. In other words, this query exposes many of the schema integration problems that the WWT will face.

More generally, building the WWT web service will require an objectified definition of Astronomy objects. It will define a set of classes, and the methods on those classes. Each archive then becomes a web service that instantiates these classes. This is a fascinating challenge for both astronomers and for computer scientists.

## Summary

The primary goal of the World Wide Telescope is to make Astronomers more productive and to allow them to better understand their data. But, it is also an archetype for the evolution of Computational Science from its simulation roots to the broader field of capturing, organizing, analyzing, exploring, and visualizing scientific data. The World Wide Telescope is a prototype for this new role, but similar trends are happening in genomics, in economics, in ecology, and in most other sciences.

This transformation poses interesting challenges to the database community which will have to deal with huge datasets, with richer datatypes, and with much more complex queries. Federating the archives is a good test of our distributed systems technologies like web services and distributed object stores. The WWT poses a challenge to the data mining community since these datasets are so huge and the data has such high dimensionality. It is an excellent place to compare and evaluate new datamining algorithms, since the data is public. It is a challenge to statisticians to develop algorithms that run fast on very large datasets. The WWT also poses the challenge of making it easy to visually explore the data, posing queries in natural ways, and seeing the answers in intuitive formats. Last but perhaps most important, the WWT can be a valuable resource to teach the new astronomy and also to teach computational science.

## Acknowledgments

Jordon Raddick led the SkyServer education effort. Tom Barclay, Tamas Budavari, Tanu Malik, Peter Kunszt, Don Slutz, Jan Vandenberg, Chris Stoughton, and Ani Thakar, helped build the SkyServer and SkyQuery. HP, Microsoft, and Fermilab support the SkyServer. Roy Williams, Julian Bunn, and George Djorgovski are building VirtualSky.